# Detailing secondary frontal bore of internal tides breaking above deep-ocean topography

**by Hans van Haren**

Royal Netherlands Institute for Sea Research (NIOZ), P.O. Box 59, 1790 AB Den Burg, the Netherlands.
e-mail: hans.van.haren@nioz.nl

**Abstract**

Above steep deep-sea topography internal tidal waves may break vigorously. The associated turbulent mixing is important for resuspending matter, bringing it tens of meters away from the seafloor for redistribution. While intense turbulence-generation occurs around a primary (frontal) bore during each transition from warming downslope to cooling upslope phase of the internal (tidal) carrier wave, a secondary bore can appear about half a wave-period later before the turn to the warming phase. As will be demonstrated from a 100-day mooring array consisting of 200 high-resolution temperature sensors between h = 6-404 m above a steep slope of a large North-Atlantic seamount and a low-resolution acoustic Doppler current profiler sampling between 50 and 450 m, secondary bores show about the same turbulence intensity as around primary bores but they generally show larger overturns that always reach the seafloor. The secondary bores associate with a sudden drop in along-isobath flow speed, a (renewed) increase in upslope flow of up to $|0.2|$ m s$^{-1}$, and with first-harmonic quarter-diurnal periodicity which provides a spectral peak for turbulence dissipation rate. While each bore is different in appearance, varying from curved like a primary bore to almost straight upward with a ragged bore, secondary bores are in a first approximation forward breaking in contrast with backward breaking primary bores.

# 1 Introduction

Since Munk's (1966) suggestion, with follow-ups by, e.g., Armi (1978; 1979), Eriksen (1982), Thorpe (1987), Garrett (1990), Munk and Wunsch (1998) and Sarkar and Scotti (2017), most deep-ocean mixing seems to occur in the vicinity of sloping bottom topography, notably via the breaking of internal waves. Given the omnipresent stable vertical density stratification in the ocean interior, internal waves are present virtually everywhere. As their dominant sources are tidal motions and passing atmospheric disturbances or fronts on the rotating Earth, internal wave breaking is expected to have the same periodicity as its source. Given the strong nonlinear character of wave-breaking however, tidal periodicity may not necessarily be dominant for every parameter. For example, multiple bursts of wave-breaking have been observed during a tidal period above Great Meteor Seamount (van Haren and Gostiaux 2012) and in the thalweg of Whittard Canyon (van Haren et al. 2022). This leads to investigation of the non-tidal periodic wave-breaking above sloping topography.

Steep topography sees a (primary) frontal bore (henceforth bore in short) propagating up its slope leading every transition from warming to cooling tidal phase (Klymak and Moum 2003) with associated strong upward flows and sediment resuspension (Hosegood et al. 2004). Further detailing observations have demonstrated intense turbulence just preceding the bore but not touching the seafloor, and, about 2 h later during the upslope flow, a secondary bore (van Haren and Gostiaux 2012; van Haren et al. 2022). Whilst the primary bore has been extensively studied demonstrating shapes varying from rounded 'dust-storm-frontal' to near-vertical ragged shapes (van Haren 2006), the secondary bore has barely been studied also because it is generally weaker in horizontal density gradient (van Haren and Gostiaux 2012). However, despite its appearance in weaker stratified waters, it may grow considerably taller than a primary bore and, importantly, it does reach the seafloor to within the lowest few meters of the water column. It may thus have an impact on sediment resuspension from the seafloor up into the ocean interior and is worth a detailed investigation.

In this paper, the secondary bore is studied in some detail using moored high-resolution temperature (T)-sensor and acoustic Doppler current profiler (ADCP) observations above a steep slope of Josephine seamount in the Northeast Atlantic Ocean. The slope is approximately planar, not incised by canyons, so rather 2D than complex 3D. While the secondary bore above Great Meteor Seamount occurred 2-3 h in the middle of the upslope tidal phase (van Haren and Gostiaux 2012; van Haren et al. 2022), the observations above Josephine demonstrate the secondary bore later in the upslope phase (van Haren 2017). Thereby above Josephine it contributes to quasi quarter-diurnal first-harmonic periodicity, but, akin to variable band-broadening arrival times of the primary bore (van Haren 2006) it will be demonstrated that characteristics of every secondary bore differ in timing, shape, strength, turbulent overturning and, potentially, environmental effects. Like the primary bore, the secondary bore is crucial in turbulence close (to within a few meters) from the seafloor, in which time-averaged vertical diapycnal turbulent diffusivity $K_z$ increases towards to the seafloor (verified down to 1 m from the seafloor in van Haren and Gostiaux 2012). This implies the importance of restratification due to internal wave breaking close to a sloping seafloor, as opposed to a reduction of $K_z$ during the warming phase of the tide when wave breaking does not reach the seafloor.

**2 Technical details**

A 450-m tall array was moored at 37° 00′N, 013° 47′W, 1890 m water depth on the eastern flank of Mount Josephine (Fig. 1), about 400 km southwest of Lisbon (Portugal) between 2 May (yearday 121) and mid-August 2015. A total of 200 'NIOZ4' high-resolution T-sensors were taped equidistantly at 2.0 m intervals to the nylon-coated 400-m long 0.005-m diameter steel (total 0.0063-m diameter) cable. A single large elliptic buoy was at z = -1463 m, 27 m above the uppermost T-sensor which was at h = 404 m above the seafloor. The lowest T-sensor was at h = 6 m. A 2.2-kN net buoyancy held the entire assembly tautly upright. Under maximum 0.2 m s$^{-1}$ current amplitudes, the low-drag mooring did not deflect from the vertical by more than 1°, i.e. the top-buoy moved <7 m horizontally and <0.1 m vertically. The average local bottom slope of about 10º is about twice the value ('supercritical') than the average angle of semidiurnal internal tide rays under average local vertical density stratification conditions. It is about equal ('critical') to the angle of internal waves at first-harmonic frequencies. The mooring was between 600 and 1000 m below the nearest sub-summit at 800 m water depth. Mount Josephine's absolute summit extends up to 250 m water depth. The mooring was also well below Mediterranean Sea outflow waters, generally found between 1000 and 1400 m, so that salinity-compensated apparent density inversions in temperature are expected to be minimally disturbing the records.

The top-buoy held a 'down-looking' Teledyne/RDI 75-kHz, 20°-slant angle to the vertical, four-beam ADCP, which measures currents in three directions, **U** = [u, v, w] averaged over all beams, and acoustic echo intensity in each beam, at 90 vertical bins of 5 m. The first bin was around the depth of the uppermost T-sensor. The lowest bin of good data was bin 70, at z = -1840 m, due to side-lobe reflection off the seafloor. There, the beam spread or horizontal averaging distance for current components was about 250 m. The ADCP stored data at once per 300 s, the T-sensors at once per 1 s.

NIOZ4 are self-contained T-sensors, with a sensor tip smaller than 1 mm, a precision better than $5\times10^{-4}$ °C after drift-correction, a response time of <0.5 s, a noise level of $<1\times10^{-4}$ °C, and a drift of about $1\times10^{-3}$ °C mo$^{-1}$ after aging (van Haren 2018). Every four hours, all sensors are synchronized via induction to a single standard clock, so that the entire 400-m vertical range is sampled in less than 0.02 s. Of the 200 T-sensors, 18 showed calibration, noise, battery or other electronic problems. Their data are replaced by linear interpolation between neighboring sensors during post-processing.

The T-data are subsequently converted into 'Conservative' (~potential) Temperature data $\Theta$ (IOC, SCOR, IAPSO 2010). They are used as tracer for density anomaly $\sigma_{1.7}$ variations following the relation $\delta\sigma_{1.7} = \alpha\delta\Theta$, $\alpha$ = -0.044±0.003 kg m$^{-3}$ °C$^{-1}$ (van Haren 2017), where subscript 1.7 indicates a pressure reference of 1700 dbar ($1.7\times10^7$ Pa). This relation is established from data between $-2000 < z < -1500$ m of a shipborne CTD-profile obtained within 1 km from the mooring site. Further post-processing includes bias-correction to smooth 398-m vertical profiles over 4-day data intervals (see van Haren 2018 for details). Given the low noise levels and the bias-smoothing, local 2-m small-scale buoyancy frequencies ($N_s$) are calculated as low as $N_s < f$, where f denotes the local inertial frequency or Coriolis parameter.

Thus, given the reasonably tight density-temperature relationship, the number of T-sensors and their spacing of 2.0 m, in combination with their low noise level, allows for accurately estimating turbulence parameters like dissipation rate $\varepsilon$ and $K_z$ via the reordering of unstable overturns making vertical density-profiles statically stable (Thorpe 1977). These overturns follow after reordering every 1-s the 398-m high potential density (Conservative Temperature) profile $\sigma_{1.7}(z)$, which may contain unstable inversions, into a stable monotonic profile $\sigma_{1.7}(z_s)$ without inversions. After comparing observed and reordered profiles, displacements $d = \min(|z-z_s|)\cdot\text{sgn}(z-z_s)$ are calculated necessary to generate the reordered stable profile. Certain tests apply to disregard apparent displacements associated with instrumental noise and post-calibration errors (Galbraith and Kelley 1996). Such a test-threshold is very low for NIOZ T-sensor data, $<5\times10^{-4}$°C. Otherwise, identical methodology is used as proposed in (Thorpe 1977) and used in, e.g., (Levine and Boyd 2006; Aucan et al. 2006; Nash et al. 2007), see (van Haren and Gostiaux 2012) for details. It includes a constant mixing efficiency of 0.2 (Osborn 1980; Oakey 1982), an Ozmidov $L_O$ – root-mean-square (rms) overturn scale $d_{rms} = (1/n\Sigma d^2)^{0.5} = L_T$ ratio of $L_O/L_T = 0.8$ (Dillon 1982) over many n-samples and the computation of buoyancy frequencies $N_s$, and their 100-m large-scale mean N, from the reordered stable density (temperature) profiles (Thorpe 1977; Dillon 1982). Then, using d rather than $d_{rms}$ as explained below,

$$\varepsilon = 0.64 d^2 N_s^3, \qquad (1)$$

$$K_z = 0.2\varepsilon N_s^{-2}. \qquad (2)$$

In (1), and thus (2), individual d replace their rms-value across a single overturn as originally proposed by Thorpe (1977). The reason is that individual overturns cannot easily be distinguished, because overturns are found at various scales with small ones overprinting larger overturns, as one expects from turbulence.

Instead, non-averaged d are used in (1) for high-resolution time-vertical images of $\varepsilon(t, z)$, the focus here. Subsequently, 'mean' turbulence values are calculated by arithmetic averaging over the vertical <...> or over time [..], or both, which is possible using moored high-resolution T-sensors (van Haren 2017). This ensures the sufficient averaging required to use the above mean constant values (e.g., Osborn 1980; Oakey 1982; Gregg et al. 2018).

Using similar T-sensor data from Great Meteor Seamount, van Haren and Gostiaux (2012) found turbulence values to within a factor of three similar to those inferred from ship-borne CTD/LADCP profiler data using a shear/strain parameterisation near the seafloor. Their values compare well with microstructure profiler-based estimates in similar sloping topography areas by Klymak et al. (2008). Comparison between calculated turbulence values using shear measurements and using overturning scales with $L_O/d_{rms} = 0.8$ from areas with such mixtures of turbulence development above sloping topography led to consistent results (Nash et al. 2007), after suitable averaging over the buoyancy scales. Thus, from the argumentation above and the reasoning in Mater et al. (2015), internal wave breaking unlikely biases turbulence dissipation rates computed from overturning scales by more than a factor of two to three, provided some suitable time-space averaging is done instead of considering single profiles. This is within the range of error.

**3. Observations**

First-month overview-data (van Haren 2017; vH17 for short) demonstrate dominant semidiurnal tidal currents and temperature variations over the entire 400-m range of observations (Fig. 3 in vH17), with spectral peaks also at first-harmonic frequencies, and also from turbulence dissipation rate series (Fig. 4 in vH17). During springtide in currents, e.g. on days 128 and 142, the isotherm-excursions exceed a vertical range of 200 m crest-trough (Figs 3,5 in vH17), like above Hawaiian Ridge (Levine and Boyd 2006; Aucan et al. 2006). Temperature seems to vary regularly tidally, with warmer waters above cooler providing stable density stratification. However, the somewhat irregular tidal motions do not represent linear internal waves, but highly nonlinear ones with occasional turbulent overturning providing unclear spring-neap cycle. This contrasts with the findings above Hawaiian Ridge where also no first-harmonic periodicity was observed (Levine and Boyd 2006).

To commence the evaluation of secondary bores above Mount Josephine, a comparison is shown between primary and secondary bores from a single tidal period near spring-tide (Fig. 2). In this and subsequent figures the vertical axis indicates h (m) above the seafloor. The primary bore is seen to be in more stratified waters and extending less high above the seafloor around the time of turning from warming to cooling phase of the tide. In contrast, the secondary bore is seen to be more upright, extending O(100) m above the seafloor in much weaker stratified waters. Further details are given below.

**3.1 Secondary bores around springtide**

In a two-day detail just before current-springtide, the dominant semidiurnal and first-harmonic variations are seen in various observables (Fig. 3). Recall that the ADCP did not sample the lower 50 m above the seafloor due to sidelobe contamination, and the T-sensors not the lower 6 m above the seafloor because of the anchor-weight and acoustic releases set-up.

The semidiurnal variability appears dominant in temperature variations with isotherm excursions exceeding 100 m amplitudes especially for h > 200 m (Fig. 3a), and along-isobath current component appearing as 400-m tall columns (Fig. 3c). To a lesser degree, semidiurnal periodicities are seen in cross-isobath current component which shows considerable vertical structure (Fig. 3b) and acoustic echo amplitude (Fig. 3e). The vertical current component does not show a distinguished semidiurnal periodicity (Fig. 3d). All variables demonstrate variations to the semidiurnal variability that differ with distance from the seafloor, and which cause the time of arrival of peaks and troughs of the semidiurnal waves to appear intermittently with time. About hourly shorter periodic motions are especially noticed in temperature and vertical current component. These associate with the mean N dominating high-frequency internal waves (van Haren 2017) and which have periods of about $T_N = 1\pm0.2$ h.

Especially in h < 200 m, the semidiurnal periodic isotherm appears highly skewed as in a sawtooth, with long rising and rapid falling. The crest is displaced later in time with respect of those higher-up, which suggests upward energy propagation (LeBlond and Mysak 1978). First-harmonic periodicity is visible in temperature in cold water appearing and disappearing, in the cross-isobath current component noticeable in upslope peaks, and in echo amplitude. From the transition between warming (downslope) to cooling (upslope) phase, the primary bore leads the first first-harmonic. This first-harmonic motion starting with the most intense bore is in anti-phase with the semidiurnal motion higher-up. The following secondary bore leads the second first-harmonic motion that is roughly in-phase with the semidiurnal motion.

Although information is lacking on the full 3D-evolution of the internal waves and on the propagation of internal waves further into the interior, it seems unlikely that the first-harmonic motions are locally generated upon reflection of the semidiurnal tides, as followed from modelling studies over a 1D-slope and, e.g., over a pycnocline (e.g., Peacock and Tabaei 2005; Diamessis et al. 2014). As free internal waves at such different frequencies rapidly disperse away from their source, the present observations likely reflect nonlinear motions of harmonics that are phase-locked to the main, semidiurnal wave.

Magnification of the lower 300 m above the seafloor shows a 100-m contrast in vertical elevation between the primary and secondary bores (Fig. 4a). As primary bores are at the very edge of substantial (re)stratification in general and in thin layers, secondary bores consistently develop in weaker stratified waters (Fig. 4b). The propagation of the semidiurnal main carrier wave generated the weaker stratification. Thus, while it is noted that primary bores demonstrate generally intense turbulent mixing reaching the seafloor (Hosegood et al. 2004), secondary bores are characterized by almost as intense and >100 m tall turbulent overturns (Fig. 4c). Turbulence dissipation rate values, averaged over 6 < h < 204 m, shows variations over four orders of magnitude, albeit with a roughly dominant first-harmonic periodicity (Fig. 4d). Secondary and primary bores share most intense turbulence dissipation rates per semidiurnal period. The observations indicate higher turbulence values reaching (down to 6 m from) the seafloor more consistently for secondary than for primary bores (Fig. 4e).

Recall that the primary bores are generally found backwards breaking (van Haren and Gostiaux 2012), with a downward motion just preceding them. In contrast, the forward breaking secondary bores may have different effects on sweeping resuspended materials into the interior, resulting in less intense relative echo amplitudes found associated with these bores compared to those of primary bores (Fig. 3e). On the large semidiurnal scales, the secondary bores occur during upward motions (Fig. 3d), which are however reduced in value at the time of frontal passage. Recall the coarse spatial sampling of the ADCP, which has horizontal foot-print of about 250 m, well exceeding the turbulent overturn scales, at h = 50 m.

Details of secondary bores are presented in 5000-s long and h < 200-m vertical images (Fig. 5). The bores have a rather coarse irregular character, are about 100 m tall, and are vertically shaped straight as well as curved. The largest overturn of warm (cold) water is always displaced for-(back)ward in time and down-(up)ward, as in forward breaking wave upon shoaling, see the arrows in Fig. 5. Their appearance deviates by about 0.0226 day (one standard deviation) from the semidiurnal lunar $M_2$ period, which confirms the canonical bandwidth of 10% of the central frequency of semidiurnal internal tides that leads to intermittency, as shown for open-ocean observations (e.g., Fu 1981; van Haren 2004). In the time domain, the 10% frequency bandwidth corresponds with 5-7 days intermittently occurring internal tidal peaks (and troughs). Each bore has a different intensity of turbulence dissipation and local stratification rates, depending on the rate of preconditioning by down-welling warm waters due to some straining effect and the upslope coming cooler waters. The latter appear to move upward, to a layer between 50 and 100 m above the seafloor, which allows for or is forced by warmer waters moving underneath them down the seafloor and up, as in a forward breaking wave. This effectuates the principally shear-induced turbulent overturns reaching the seafloor, in a convective manner. The frontal 'width' in time of a temperature difference of 0.1°C varies between 0.0015 and 0.01 days (150 and 900 s), which is at least one order of magnitude longer than primary bores observed below the top of Great Meteor Seamount (van Haren and Gostiaux 2012). Fig. 6 provides a conceptual cartoon of the secondary bore.

**3.2 Other secondary bores throughout the record**

A choice of different secondary bores from the three-month mooring period (Fig. 7) shows somewhat different forms, in more slantwise verticality (Fig. 7a,c) and apparent convection from above (Fig. 7b), but otherwise similar characteristics as during springtide, including a large initial overturn of downward motion that moves warmer waters near the seafloor and up in the cooler waters behind the bore (Fig. 7d). Such overturns may be difficult to trace visually due to the smaller scale overturns overprinting the large ones.

**3.3 No secondary bores during a short period of convection burst around turn of tide**

An exceptional portion of the record occurred near a neap-tide, during which large overturning was observed but rather unclear secondary bores (Fig. 8). This period lacks distinctive first-harmonic periodicity in temperature, cross-slope flow, although some is visible in echo amplitude. The horizontal current components show a more irregular pattern in the vertical than during spring-tide. However, most different is the replacement of secondary bore occurrence prior to the turn from cooling to warming semidiurnal phase by a quasi-symmetric large instability formation on both sides of that turn of phases. In comparison with especially lower isotherms in Fig. 4, the isotherms in Fig. 8 are more symmetric around their highest point. The quasi-symmetry may be associated by the sloshing back and forth of the tide, advecting the same structure passed the T-sensors including development of the turbulent overturning in the mean-time, but we lack information of the 3D development of the internal waves. It is obvious from Fig. 8 that isotherms are less nonlinear than in Fig. 4 including a lack of lower isotherm semidiurnal saw-tooth appearance, and bore development is weaker. Apparently, this does not preclude the generation of >100 m tall overturns that provide the same mean turbulence dissipation rate within a factor of two (the standard error) compared with the secondary-bores-dominated spring-tide period of Figs 3,4.

However, their mean dissipation rate profiles differ greatly. Whilst the two-day mean of the spring-tide is only weakly decreasing with height above the seafloor (Fig. 9a), that of neap-tide decreases by almost three orders of magnitude (Fig. 9b). The latter shows larger values in the lower 150 m above the seafloor, which are predominantly attributable to the semidiurnal periodic overturning during the cooling to warming phase transition. Both profiles show a significant increase in the lower few T-sensors, from about $h = 15$ to 6 m.

Such local increase near the seafloor is observed consistently during the spring-tide secondary bores of Fig. 5 (Fig. 9a). Depending on the particular secondary bore, a 200-m tall high dissipation rate is observed, or a more variable and on average slightly weaker one, with minimum values between $50 < h < 100$ m. The 5000-s mean secondary bore profiles of Fig. 9a mimic best the two-day mean profile of Fig. 9b.

In Fig. 9b, 5000-s mean profiles of their associated primary bores are given, for comparison. These show coarsely more vertically consistent profiles, with decrease of 1-2 orders of magnitude over the 400 m range. Near the seafloor, these profiles show the largest variability possible, varying from highly increased value by more than one order of magnitude, via a half-order of magnitude increase as in Fig. 9a, a nearly constant value compared to values higher up, to a one-order of magnitude decrease. This 2.5-order of magnitude variation in the lowest T-sensors is attributable to the complex variability of primary bore development, its perpendicular or oblique propagation up a slope (Fig. 16 in Hosegood and van Haren 2004) which may result, as in this case, in the actual overturning reaching the seafloor or not. The more perpendicular a primary bore moves up isobaths, the steeper slope it senses with generally increased nonlinearity and associated wave breaking.

The 3D nature of frontal curvature will matter, but is unknown from the present 1D measurements.

## 4. Discussion

The consistency in secondary bore turbulence reaching towards the seafloor with highest dissipation rates at the lowest T-sensor level confirms previous turbulence diffusivity profiles down to 1 m above a sloping seafloor of Great Meteor Seamount (van Haren and Gostiaux 2012). This consistency contrasts with the turbulence effects on sediment resuspension of primary bores which may be one-order of magnitude more intense, but also less intense by the same amount. Primary bores are thus more variable, presumably due to their above-mentioned 3D development or precise interaction (direction) of the carrier wave with the sloping topography (Hosegood and van Haren 2004). Hypothesizing, secondary bores are less dependent on such development.

This lesser dependency of propagation direction of secondary fronts may be due to their appearance towards the end of the upslope cooling phase of the tide, rather than being dependent on the capturing of primary bore development at the turn of warming to cooling phase. Unclear is how it may also cause forward breaking as in surface waves beaching, which contrasts secondary bores from backward breaking primary bores. This may be associated with the weaker stratified pre-conditioning in which higher frequency internal waves from over-tides to the buoyancy frequency play a role and which allows for forward breaking upon shoaling. The pre-conditioning possibly by some straining effect of depressing weakly stratified warmer waters towards the seafloor seems imperative for secondary bore development.

While each bore is different in appearance, varying from curved like a primary bore to almost straight upward with a ragged bore, monthly mean spectra provide a first-harmonic peak in turbulence dissipation rate.

The vertical range of h < 100150 m of high turbulence levels due to internal waves breaking confirms high-resolution numerical modeling (Winters 2015) of turbulence generation via internal wave sloshing and breaking over a slope. Even at h = 400 m, turbulence is at least one order of magnitude more intense than in the open ocean-interior (Gregg 1989).

The present observations demonstrate that local nonlinear internal wave interaction leads to secondary bores that are important for periodic high turbulent mixing above steep deep-ocean topography. The associated continuous increase of turbulence dissipation rate, and diffusivity, with depth leaves only a very thin, less than a few meters, layer above the seafloor in which turbulence may potentially be reduced. Thus, the requirement for upwelling along sloping boundaries to compensate for downwelling in the ocean interior following turbulent mixing as suggested by Ferrari et al. (2016) may be restricted to a very limited layer above sloping seafloors. It may be enhanced following 3D aspects like due to canyons, other (less steep) topography, and during the warming phase of tidal waves.

**Acknowledgements** I thank the captain and crew of the R/V Pelagia and NIOZ-NMF for their very helpful assistance during deployment and recovery. M. Stastna (Univ. Waterloo, Canada) provided the darkjet colourmap suited for high-resolution moored T-sensor data. NIOZ temperature sensors have been funded in part by NWO, the Netherlands organization for the advancement of science.

**References**

Armi L (1978) Some evidence for boundary mixing in the deep ocean. J Geophys Res 83:1971-1979

Armi L (1979) Effects of variations in eddy diffusivity on property distributions in the oceans. J Mar Res 37:515-530

Aucan J, Merrifield MA, Luther DS, Flament P (2006) Tidal mixing events on the deep flanks of Kaena Ridge, Hawaii. J Phys Oceanogr 36:1202-1219

Diamessis PJ, Wunsch S, Delwiche I, Richter MP (2014) Nonlinear generation of harmonics through the interaction of an internal wave beam with a model oceanic pycnocline. Dyn Atmos Oc 66:110-137

Dillon TM (1982) Vertical overturns: a comparison of Thorpe and Ozmidov length scales. J Geophys Res 87:9601-9613

Eriksen CC (1982) Observations of internal wave reflection off sloping bottoms. J Geophys Res 87:525-538

Ferrari R, Mashayek A, McDougall TJ, Nikurashin M, Campin J-M (2016) Turning ocean mixing upside down. J Phys Oceanogr 46:2239-2261

Fu L-L (1981) Observations and models of inertial waves in the deep ocean. Rev Geophys Space Phys 19:141-170

Galbraith PS, Kelley DE (1996) Identifying overturns in CTD profiles. J Atmos Ocean Technol 13:688-702

Garrett C (1990) The role of secondary circulation in boundary mixing. J Geophys Res 95:989-993

Gregg MC (1989) Scaling turbulent dissipation in the thermocline. J Geophys Res 94:9686-9698

Gregg MC, D'Asaro EA, Riley JJ, Kunze E (2018) Mixing efficiency in the ocean. Ann Rev Mar Sci 10:443-473
15

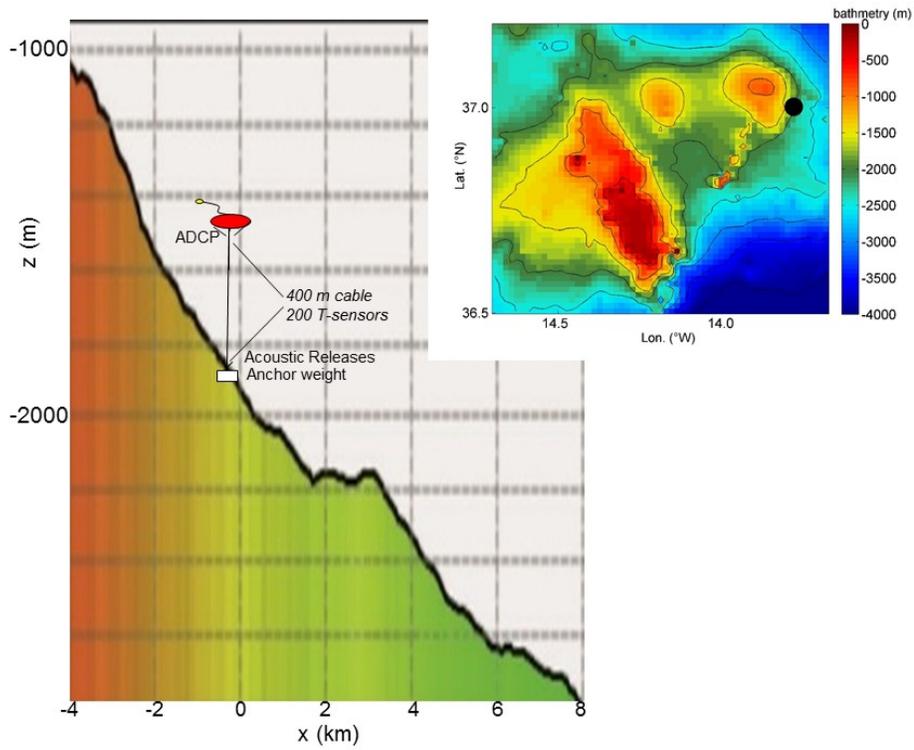

**Fig. 1** Mooring array above eastern slope of sub-summit of Mount Josephine, NE-Atlantic Ocean. The mooring sketch is above topography slope from Multibeam cross-section along 37°N. The insert bathymetry is from 1′-version of Smith and Sandwell (1997) with the black dot indicating the mooring location.

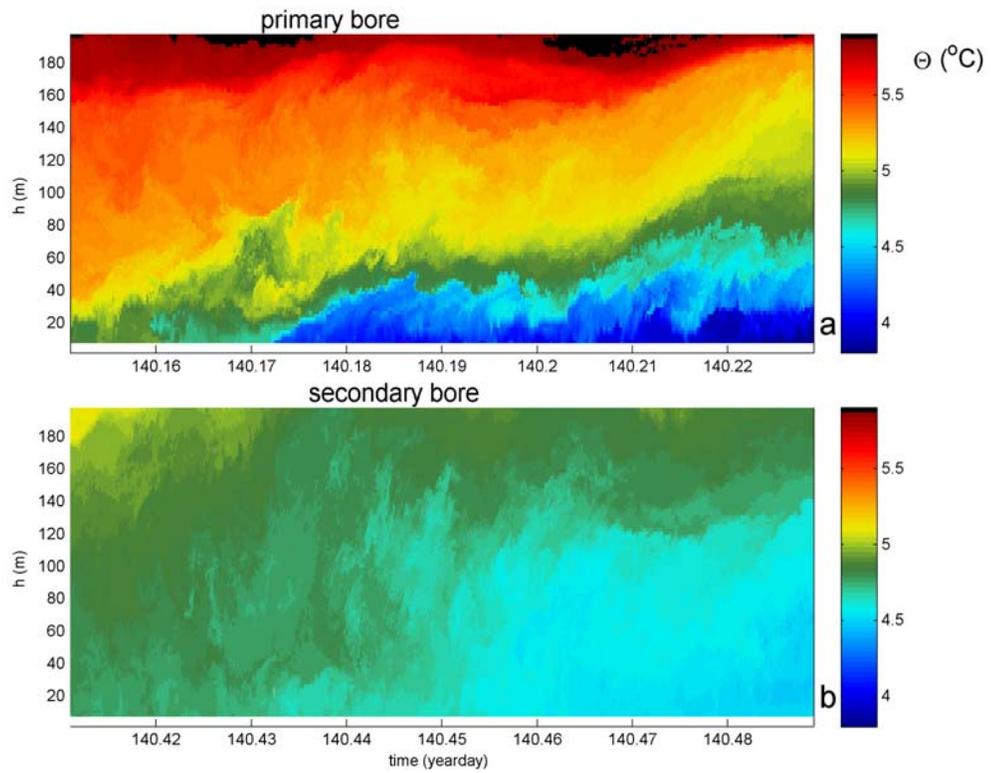

**Fig. 2** Demonstration of characteristic differences between primary and secondary bores over the sloping seafloor. Examples of 0.2-d, 200-m time-height plots of Conservative Temperature from a single tidal period. In both panels, the horizontal axis is at the level of the seafloor h = 0. **a** Primary bore. **b** Secondary bore 0.26 d later than a. The colour-scale is identical to that in a.

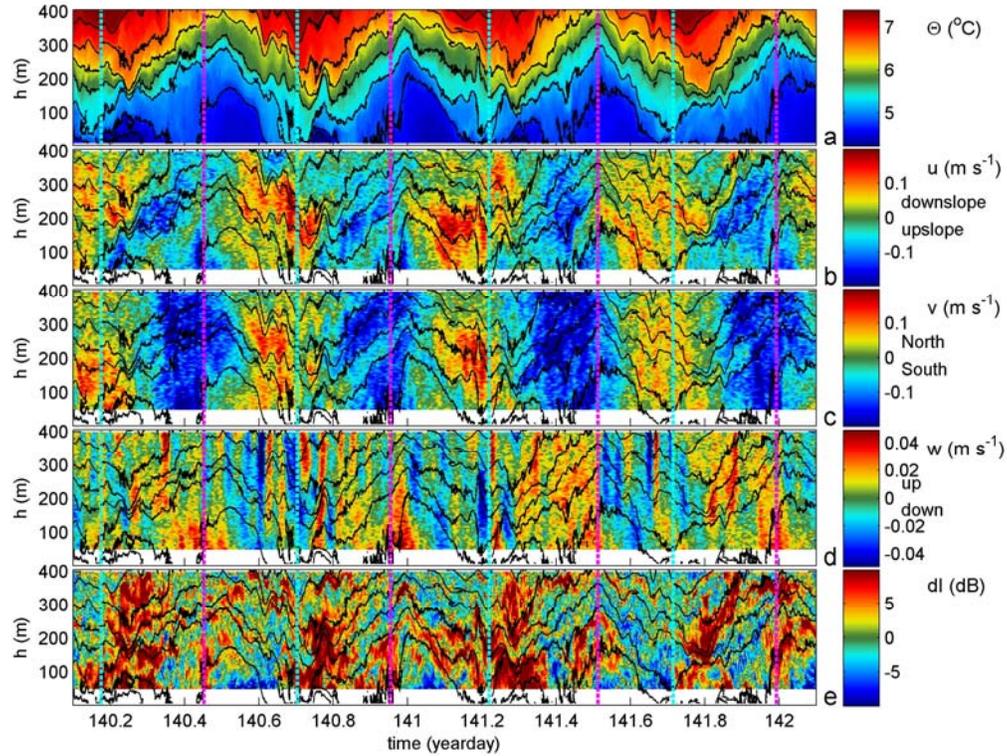

**Fig. 3** Two-day 400-m detail of moored temperature, current and acoustic echo measurements above a slope of Mount Josephine. In each panel, the horizontal axis is at the level of the seafloor. Vertical dashed lines indicate primary (light-blue) and secondary bores (magenta), approximately. **a** Conservative Temperature from high-resolution T-sensors, sub-sampled every 15 s. Black contours are drawn every 0.5°C. **b** Cross-isobath current component from ADCP, low-pass filtered at once per hour to remove noise. The black contours are copied from a. for reference. **c** As b., but for along-isobath current component. **d** As b., but for vertical current component. Note the different scale. **e** As b., but for acoustic echo amplitude relative to time-mean values.

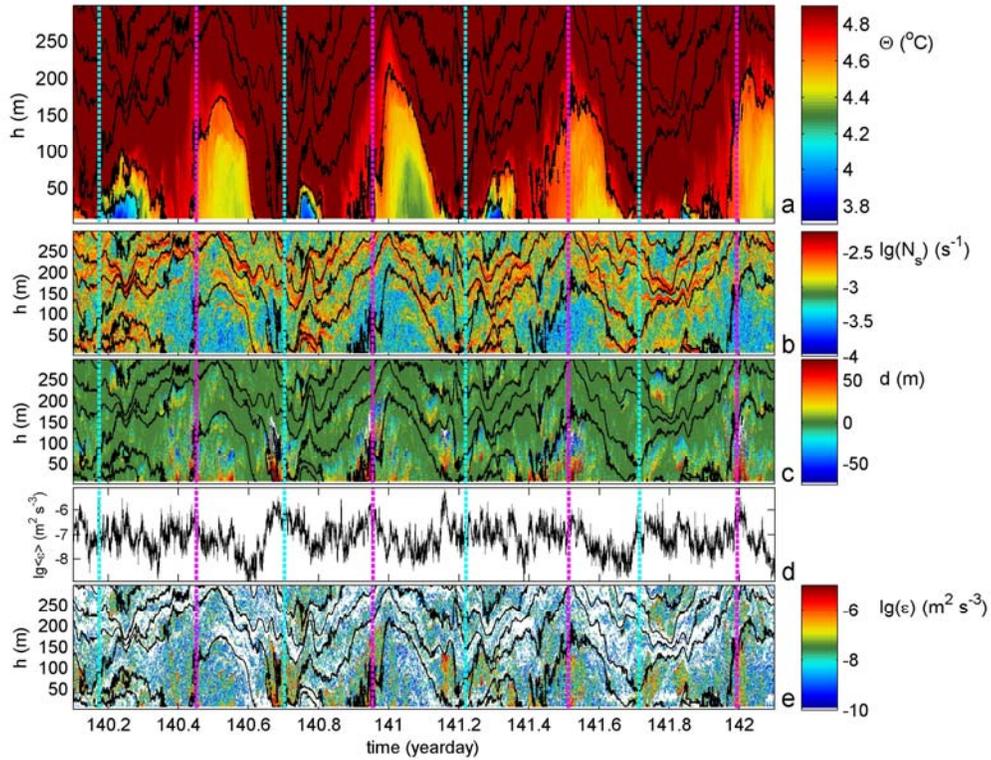

**Fig. 4** As Fig. 3, but for 300-m tall magnification of temperature and calculated turbulence values. **a** Conservative Temperature with scale modified to high-light values near the seafloor. **b** Logarithm of small-scale buoyancy frequency calculated from reordered profiles of a. **c** Displacements between original and reordered temperature profiles. **d** Logarithm of turbulence dissipation rate averaged over the lower 200 m of T-sensor range above the seafloor. **e** Logarithm of non-averaged turbulence dissipation rate.

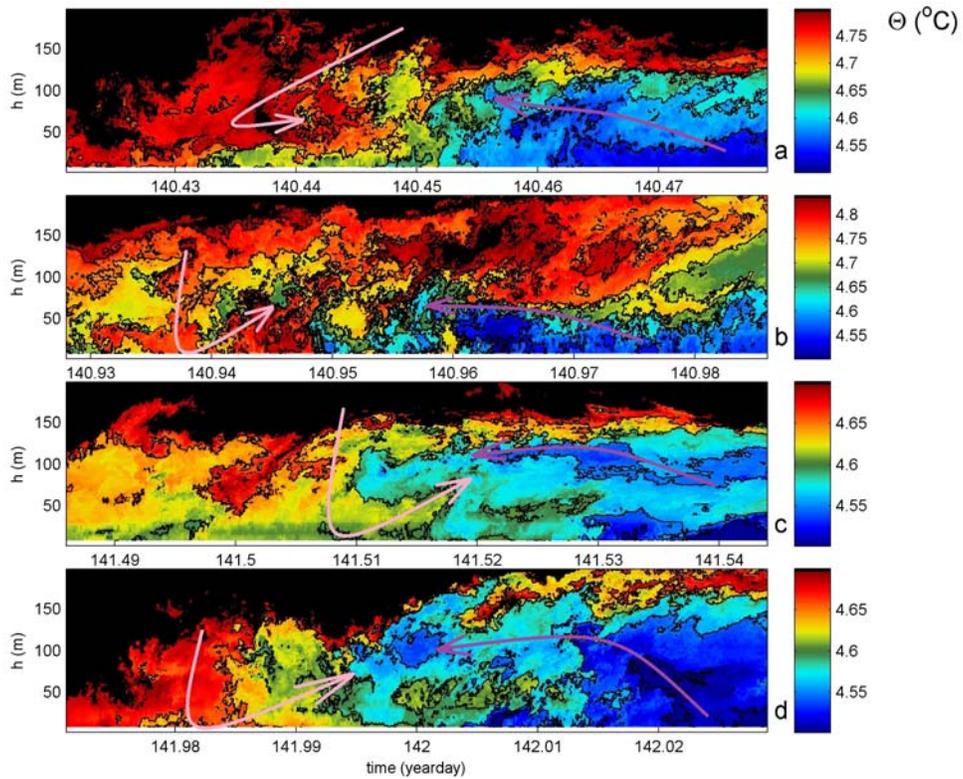

**Fig. 5** Magnifications of 5000 s and lower 200 m t-z-images of 1-s sampled Conservative Temperature around four secondary bores from the two-day period of Fig. 3a. The pink and purple arrows indicate large relatively warm and cool water displacements in each panel. Black contours are drawn 0.05°C. Note the slight differences in colour range between the panels (except c. and d.).

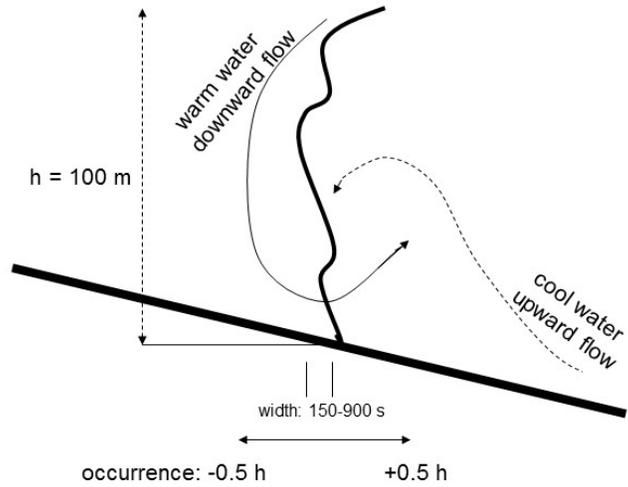

**Fig. 6** Cartoon of some characteristics of secondary bores, occurring ±0.5 h every semidiurnal tidal cycle, as discussed in the last paragraph of Section 3.1.

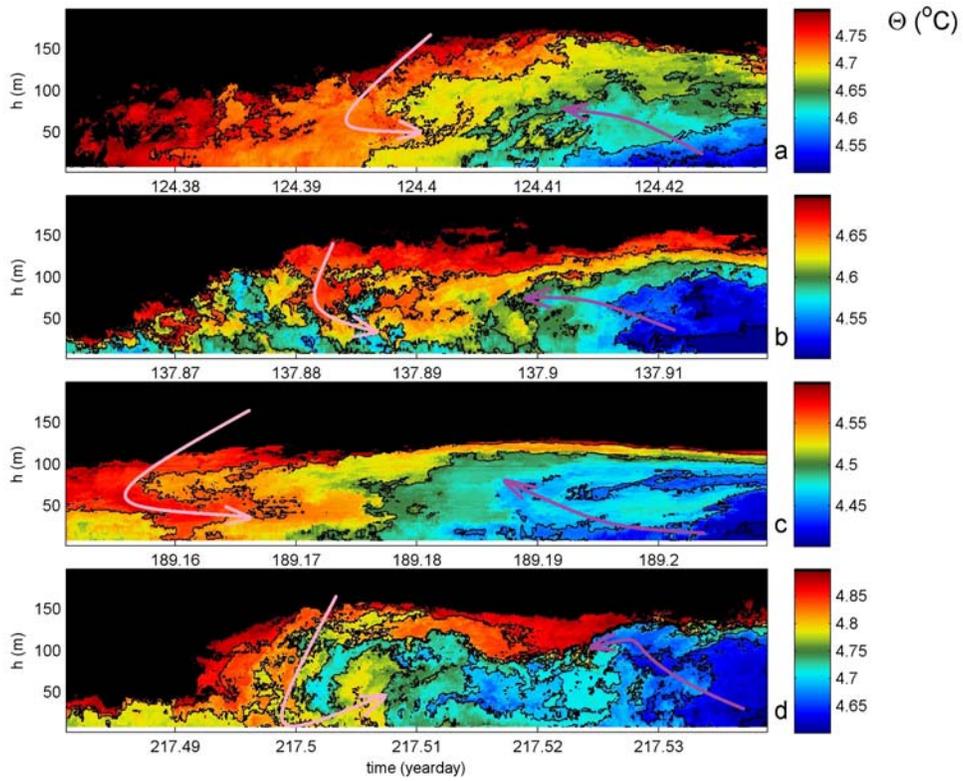

**Fig. 7** As Fig. 5, but for four different secondary bores chosen throughout the 100-day mooring period.

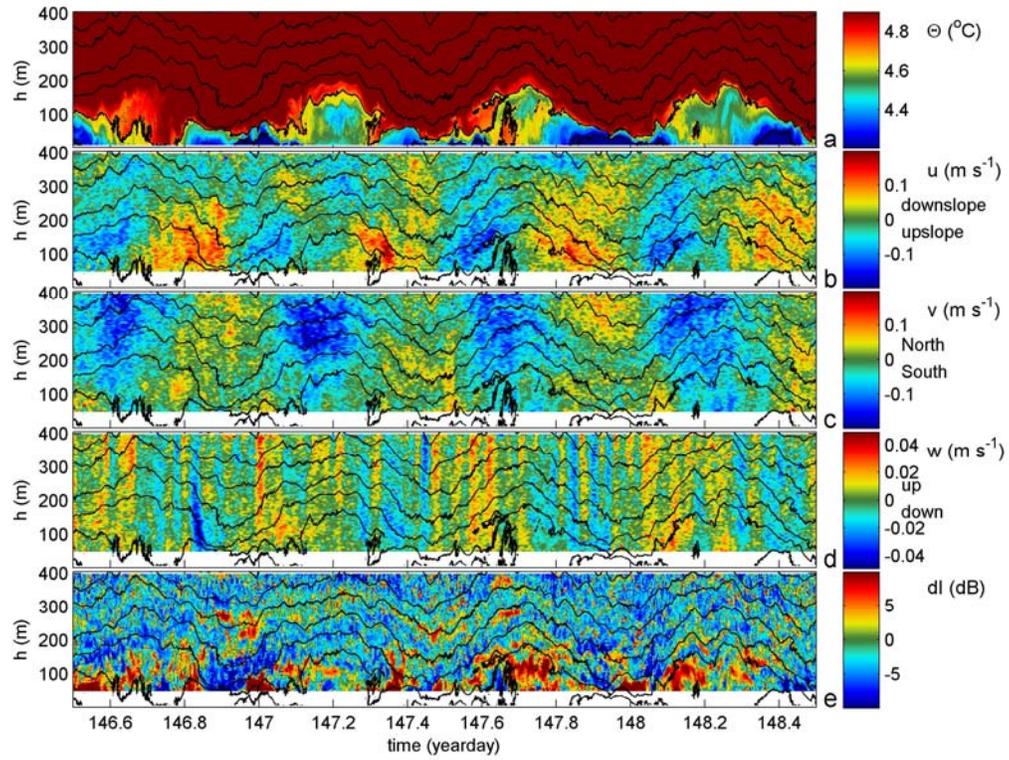

**Fig. 8** As Fig. 3, but for a rare two-day period near a local neap tide lacking secondary bores and associated upslope flow, but including strong turbulent overturning around the turn from cooling to warming tidal phase.

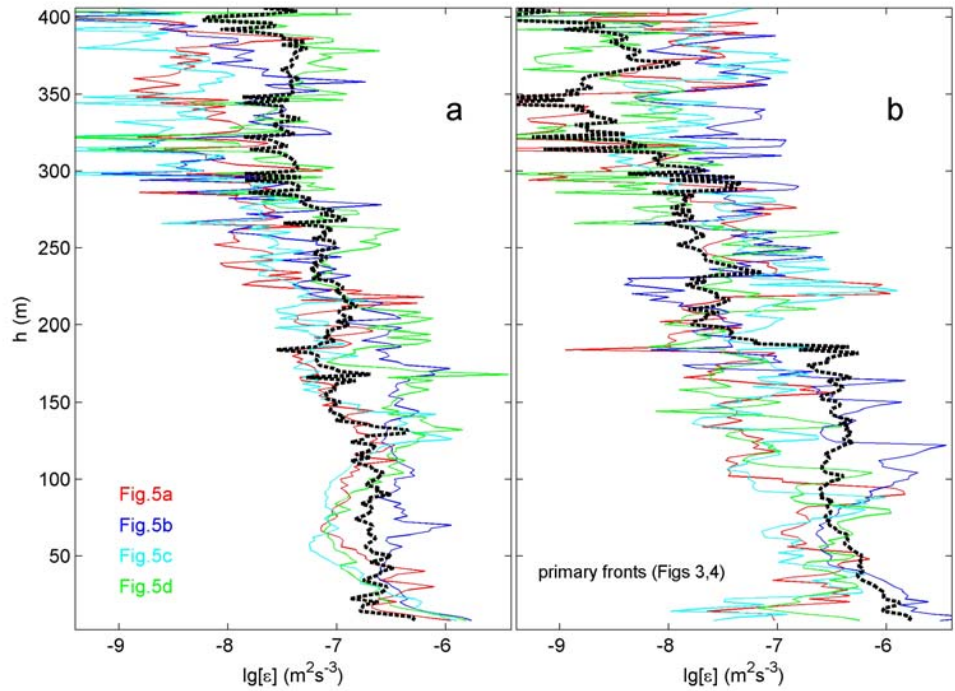

**Fig. 9** 5000-s Time-averaged turbulence dissipation rate profiles, with the seafloor at the level of the horizontal axis. **a** For data from Fig. 5. The black-dashed profile is the two-day average for the period of Figs 3,4. **b** For data around primary bores from Figs 3,4. The black-dashed profile is the two-day average for data from Fig. 8.